\documentclass[conference, letterpaper]{IEEEtran}

\usepackage{times}
\usepackage{helvet}
\usepackage{courier}

\usepackage[T1]{fontenc}
\usepackage[latin9]{inputenc}
\usepackage{color}
\usepackage{url}
\usepackage{amsthm}
\usepackage{amsmath}
\usepackage{amssymb}
\usepackage{graphicx,threeparttable}
\usepackage{footnote,epstopdf,nameref,xspace}

\usepackage{algorithm}
\usepackage{algorithmic}

\usepackage{float}
\usepackage{graphicx}
\usepackage{epsfig}
\usepackage{balance}
\usepackage{multirow}

\graphicspath{{pix/}}
\usepackage{caption}
\usepackage{subcaption}

\usepackage{listings}
\usepackage{xcolor}
\usepackage{color}
\definecolor{keywordcolor}{rgb}{0.8,0.1,0.5}
\definecolor{webgreen}{rgb}{0,.5,0}
\begin{document}

\title{On Diffusion-restricted Social Network: \\ A Measurement Study of WeChat Moments}

\author{
Zhuqi Li$^1$, Lin Chen$^2$, Yichong Bai$^1$, Kaigui Bian$^1$, and Pan Zhou$^3$\\
$^1$School of EECS, Peking University, Beijing, China\\
$^2$Department of Electrical Engineering, Yale University, New Haven, CT, USA\\
$^3$School of EIC, Huazhong University of Science \& Technology
}

\maketitle
\begin{abstract}
WeChat is a mobile messaging application that has 549 million active users as
of Q1 2015, and ``WeChat Moments'' (WM) serves its social-networking function that
allows users to post/share links of web pages. WM differs from the other social networks as it imposes many restrictions on the information diffusion process to mitigate the information overload. 
In this paper, we conduct a measurement study on information diffusion in the WM network by crawling and analyzing the spreading statistics of more than 160,000 pages that involve approximately 40 million users. Specifically, we identify the relationship of the number of posted pages and the number of views, the diffusion path length, the similarity and distribution of users' locations as well as their connections with the GDP of the users' province.
For each individual WM page, we measure its temporal characteristics (e.g., the life time, the popularity within a time period);
for each individual user, we evaluate how many of, or how likely, one's friends will view his posted pages.
Our results will help the business to decide when and how to release the marketing pages over WM for better publicity.
\end{abstract}

\section{Introduction}
\begin{figure}[htbp]
\centering
 \begin{subfigure}[b]{1in}
    \includegraphics[width=\textwidth]{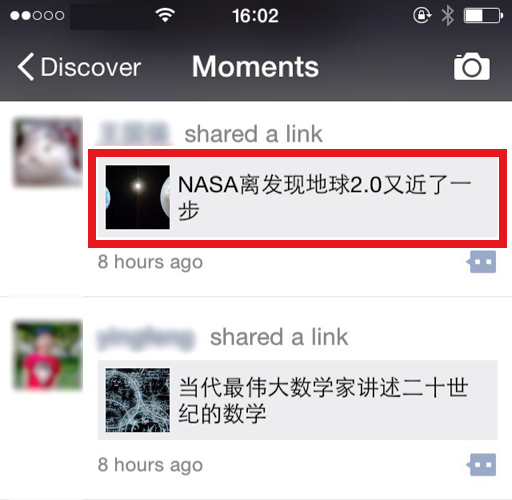}
    \caption{}
    \label{fig:like_comment}
\end{subfigure}
 \hfill
 \begin{subfigure}[b]{1in}
    \includegraphics[width=\textwidth]{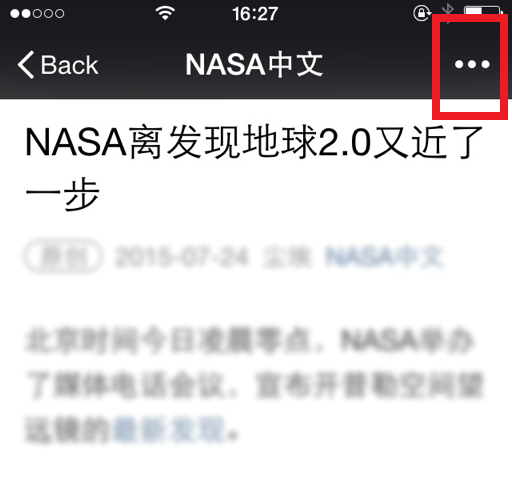}
    \caption{}
    \label{fig:noshare}
\end{subfigure}
  \hfill
 \begin{subfigure}[b]{1in}
    \includegraphics[width=\textwidth]{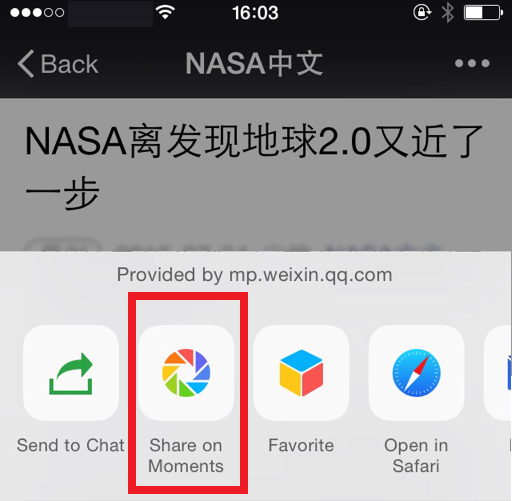}
    \caption{}
    \label{fig:share}
\end{subfigure}
\caption{WeChat Moments: (a) A posted link; (b) The menu button in the posted WM page; (c) The share button.}
\label{fig:WM}
\end{figure}

WeChat is a mobile messaging application that has attracted 549 million active users as
of Q1 2015, most of whom are from China mainland. Launched in 2012, ``WeChat Moments'' (WM) serves WeChat's social-networking function that allows users to post/share on their walls with links (Figure~\ref{fig:like_comment}) of web pages in HTML5 (or H5) (a.k.a. WM pages, see Figure~\ref{fig:noshare}). Since WeChat is the most dominant mobile social application in China~\cite{schiavenza2013wechat}, WM has been over-exploited by businesses to spread information for viral marketing/advertising. However, the information overload problem may also arise in WM, when businesses excessively circulate their advertising pages in the social network, which has undermined the user experience~\cite{ellison2007social,gilbert2013need}.

To mitigate the information overload problem, WM differs from the other social networks by imposing the following restrictions on the information diffusion process~\cite{wu2014wechat}.
\begin{itemize}
  \item \textbf{The maximum number of friends}: Each user is limited to having no more than 5,000 friends, which avoids the information overload caused by a user who has more than many followers/fans and deliver excessive information;
  \item \textbf{No access to strangers' posts}: Two users cannot see each other's posted content if they are not connected as friends. This access control tries to confine the information diffusion within a community of acquaintances;
  \item \textbf{The hidden ``share'' button for reposting}: A user has to click the upper-right menu button with an icon ``\colorbox{black}{\textcolor{white}{$\cdots$}}'' (highlighted in Figure~\ref{fig:noshare}) to repost a link (a WM page) (the share button is highlighted in Figure~\ref{fig:share}). This somewhat de-motivates users to repost a WM page.
\end{itemize}
To understand information diffusion process of an H5 page in WM, we answer three questions in this study: (1) How much will these restrictions affect/throttle the WM page diffusion? (2) What kind of temporal characteristics of WM page diffusions can be leveraged by businesses for better advertising? and (3) What are the typical features of a highly-influential users, when the number of friends is no longer a strong indicator of one's influence?

In this study, we conduct a measurement study of WeChat Moments, by crawling and analyzing the spreading statistics of more than 160,000 WM pages created by businesses in 2015 (with an average number of 10 million page views every day by 40 million users). We investigate three aspects of the information diffusion in the WM network, as will be summarized below.
\begin{itemize}
  \item  \textbf{Diffusion graph in WM}: We have examined the number of posted pages, the number of page views, the diffusion path length, users' locations, and identified the relationship among these factors or their implicit connections to macro-economic factors (e.g., GDP of a user's province).
  \item \textbf{Influence of WM pages}: We have measured the life time, the popularity, influence range of individual posted pages, from which the business can benefit in when and how to develop marketing strategies for better publicity.
  \item \textbf{User influence}: We evaluate how many of, or how likely, one's friends will view his posted pages, and our research findings indicate that the number of friends is no longer a strong indicator of one's influence. Instead, we discuss decisive features that the influential users may have, e.g., the number of identified friends, and the ability of attracting similar friends for page views.
\end{itemize}
To the best of our knowledge, this is the first measurement study on the information diffusion in WeChat Moments networks.


\section{Related Work}
In this section, we present a brief literature review on two broad categories of online social networks, namely ``open'' and ``anonymous'' social networks.

\subsection{Open Social Networks}
Over the last few years, the rising popularity of open social networks has introduced many problems for researchers. Researchers have focused their interests on analysis, measurement, and experiments of information diffusion, user behaviors, community structure, advertisement, etc. in Twitter~\cite{kwak2010twitter}, Facebook~\cite{ellison2007benefits}, Sina Weibo~\cite{gao2012comparative}~\cite{wang2015profiling}, and even technical community like Stack Overflow~\cite{anderson2012discovering}. In these open social networks, a user's post or page can be seen by anyone if no access restriction is enforced by the user. This greatly stimulates users to share and spread the information in the network, and boosts the prosperity of open social networks. However, it may lead to information overload problems as well as privacy concerns.

In the Wechat Moments, the policy of ``no access to strangers' posts'' can provide a mitigation of information overload, and a protection for users' privacy. In the meanwhile, this may arouse a new question: how much will this access restriction affect the message diffusion in Wechat Movements? This will be answered in our study.

\subsection{Anonymous Social Networks}
Anonymous social networks provide users with a platform to post messages and communicate without showing their real identity. Some researchers have focused on studying many anonymous communicating platforms such as Whispers~\cite{wang2014whispers}, Yik Yak~\cite{northcut2015dark} and YouBeMom~\cite{schoenebeck2013secret}. Unlike the open social networks, these anonymous platforms well support the protection for users' privacy, as there is no way to figure out the real identify of any user when the information diffusion between two users is established via their ``weak ties'' such as on their location, interests, or friends in common, rather than the real-world friend relationship. However, such a relationship over weak ties may discourage the users to share a post or even disrupt the diffusion process of a message.

The communication in Wechat Movements is based on strong ties (e.g., friend relationship in the real world), and WM seeks to strike a balance in the tradeoff between privacy protection and non-excessive information diffusion in social networks, which is the motivation behind our study as well.


\section{WeChat Moments (WM) Dataset}~\label{sec:dataset}
In this section, we introduce what is a WM page and how our WM data are collected.

\textbf{WM page}: In the WM network, the links shared/posted by users usually lead to web pages in HTML5 (H5). Such pages provide users with interactive operations such as an online greeting card, a lightweight online game (e.g., flappy birds), psychological test, etc. Hence, a web page that can be shared over the WM network is also called a \emph{WM page}. A WM page can be launched by the WM service provider (Tencent), or a third-party web developer.

\textbf{Page diffusion process}: Suppose that users~$i$ and $j$ are friends in the WM network. When user~$i$ shares the link of a WM page with his friends, user~$j$ may click the shared link to view the content of this page. If user~$j$ finds this page interesting, he may further repost the page link to his friends, so that more users would have the chance to view this WM page. As this process is similar to the spread of infection,
\begin{itemize}
  \item we call a user an \emph{infected} user of a WM page if he views the page;
  \item we call a user an \emph{infectious} user of a WM page if he views the page and reposts the page link.
\end{itemize}

\textbf{WM data collection}: Our goal is to collect the statistics of WM page diffusions.
We use the Application Programming Interface (API) provided by a business WM page development platform for crawling diffusion trails of pages created over the platform. Based on the collected data, we are able to construct a diffusion graph for each posted WM page. The dataset contains about 160,000 pages created by businesses from April 22th to June 5th 2015, which attracted about 40 million users and an average number of 10 million page views every day.


\textbf{Data format of page view records}: A page view record in our dataset is a 5-tuple in the following format:
\[
<U_1, U_2, PID, IP, t>,
\]
where $U_1$ is the ID of the user whose posted page gets viewed; $U_2$ is the ID of the user who views the page posted by user~$U_1$; $PID$ is the ID of the WM page that is assigned by the WM page development platform; $IP$ the IP address of the page viewer, i.e., user~$U_2$; and $t$ is the time when the page view event happens. The whole tuple records the page view event where user~$U_2$ at the address~$IP$ views page~$PID$ posted by user~$U_1$ at time $t$.

\section{Analysis of Diffusion Graphs}~\label{sec:basic}
In this section, we analyze the characteristics of the overall diffusion graph, including the diffusion path lenth, total numbers of posted pages and views, users' locations and their implicit connections to the GDP of a user's province.

\subsection{Diffusion path length}
We first consider the diffusion path length which means how many steps are needed for a page to be spread to a user from the source. Degrees of separation has become an important tool to understand the social network structure owing to the interesting ``six degrees of separation'' experiment by Stanley Milgram~\cite{milgram1967small}. The diffusion graph we constructed is different from the underlying network graph in WM, as the former only contains those edges and vertices (users) that are infected by WM pages, while the latter has all the edges and vertices in the WM network.

\begin{figure}[ht]
\begin{minipage}[t]{0.45\linewidth}%
\centering
\includegraphics[width=1.5in]{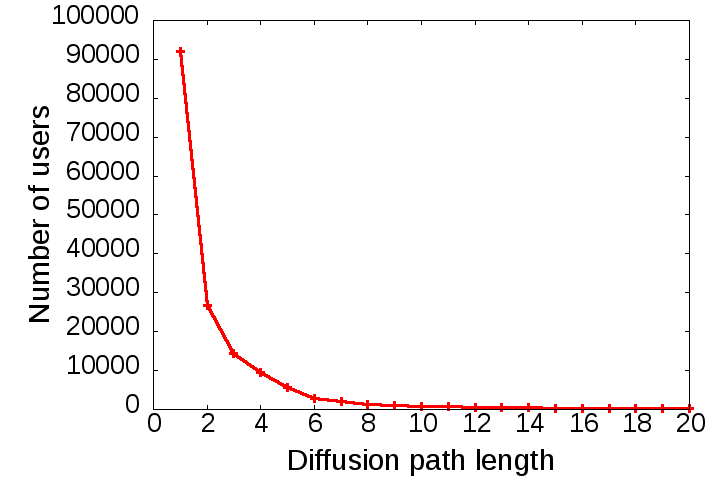}
\caption{Diffusion path length.} \label{fig:bian-path-length}
\end{minipage}
\hfill%
\begin{minipage}[t]{0.45\linewidth}%
\centering
\includegraphics[width=1.5in]{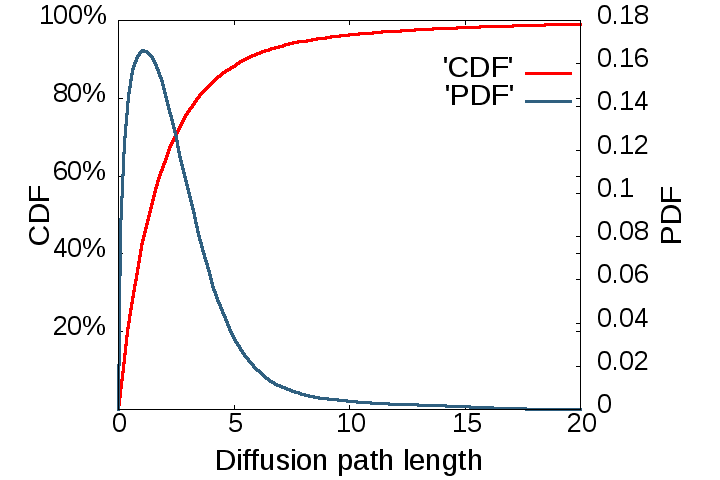}
\caption{CDF and PDF of diffusion path length.} \label{fig:bian-path-length-cdf}
\end{minipage}
\end{figure}

\begin{figure*}[ht]
 \begin{minipage}[t]{0.31\linewidth}%
\centering
\includegraphics[width=\columnwidth]{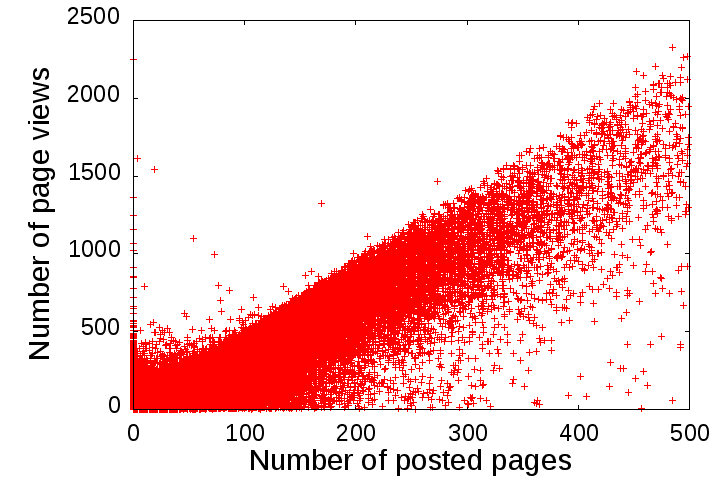}
\caption{The numbers of posted pages and page views.} \label{fig:bian-post-viewed}
\end{minipage}
\hfill%
\begin{minipage}[t]{0.31\linewidth}%
\centering
\includegraphics[width=\columnwidth]{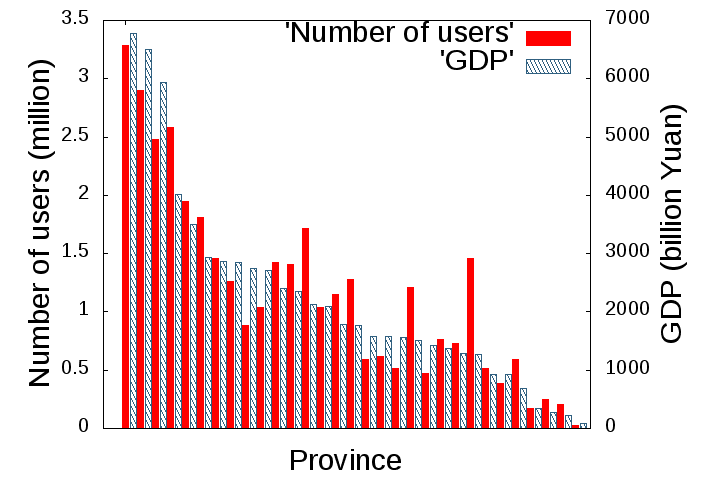}
\caption{The relationship between the number of infected WM users and the GDP of the user's province.} \label{fig:bian-loc-gdp}
\end{minipage}
\hfill%
\begin{minipage}[t]{0.31\linewidth}%
\centering
\includegraphics[width=\columnwidth]{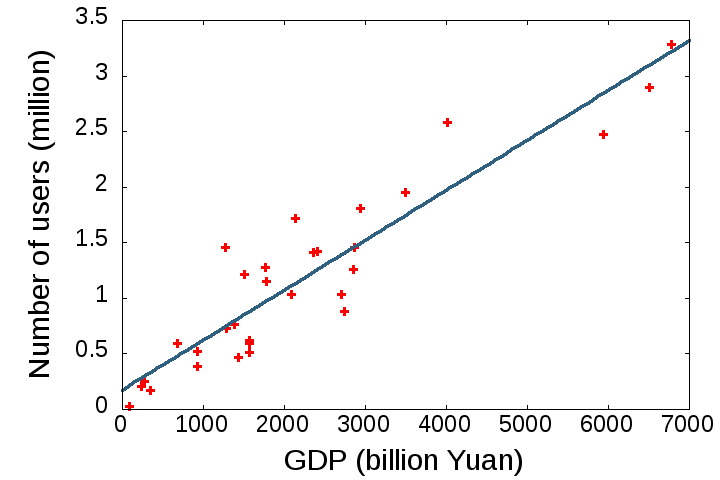}
\caption{The correlation between the number of infected WM users and the GDP of the user's province.} \label{fig:bian-usr-location}
\end{minipage}
\end{figure*}

The diffusion graph can reflect the path over which a page diffuses in the WM network. Because Similar to the sampling method used in \cite{wang2014whispers,ahn2007analysis}, we randomly choose 1,000 pages from the dataset and compute the shortest diffusion path to all the users they can infect. We plot the result in Figure~\ref{fig:bian-path-length}. About 90,000 users can directly view a chosen page by only one step. That is to say, a large portion of users view the page directly from the source. From this phenomenon, we can infer that the WM pages diffuse in the neighborhood of the source. We also notice that  few pages can have a diffusion path length longer than six, which implies that the source can reach a user by a diffusion path no longer than six, i.e., the ``six degrees of separation'' can be also applied in the diffusion graph of WM.

To take a closer look at the distribution of the number of users for a certain diffusion path length, we plot the Cumulative Distribution Function (CDF) and Probability Density Function (PDF) in Figure~\ref{fig:bian-path-length-cdf}. We observe that for 90\% users, they are infected by WM pages via a diffusion path that has a length no longer than five; there is almost no diffusion path that has a length above 15.

\textbf{Implications}: As has been mentioned before, WM has an access permission restriction that prohibits strangers to see each other's posted pages. This is different from other social networks where a user has the opportunity to see a stranger's posted information, which promotes the information diffusion between strangers. However, in the WM networks, the page is only spread between friends, or even within a neighborhood of acquaintances, which throttles most of the long diffusion paths. Hence, there is no chance for a celebrity to deliver information to his fans in WM, and marketing should focus on the acquaintance referrals.


%
%

\subsection{Numbers of posted pages and page views}
In this part, we investigate the relationship between the number of pages posted by a user and the number of views to his pages. We try to answer the question: whether a user can get a higher influence by posting more pages? The answer is ``Yes''.

We compute the average numbers of posted pages and page views for all users, and we plot them in Figure~\ref{fig:bian-post-viewed}. We find that the red points are mostly distributed in a ``belt-like'' region that can be roughly defined by two lines $y = 4.146x + 177.3$ and $y = 4.146x - 684.1$. As the number of posted pages increase, the number of received page views also grows, leading to a rising influence.

\textbf{Implications}: Here we draw an intuitive conclusion---businesses in WM can get a higher influence by posting more pages, which is the same as the marketing strategy in other social networks~\cite{krishnamurthy2008few}.


\subsection{User location and GDP}
As is known, users' location may be closely related to their real-world living conditions. With collected IP addresses, we can tell the rough location (the located province) of a user who gets infected by a WM page by using an open API that translates an IP address to a geographical location~\cite{googleapi}.


We count the number of users located in every province of China, and plot the number of users and the GDP values of all provinces of China in Figure~\ref{fig:bian-loc-gdp}\footnote{The GDP of China in year 2014 is from National Bureau of Statistics of China~\cite{gdp2014}.}. All provinces are listed over the horizontal axis in a descending order of their GDP values. Each blue bar in the figure represents the GDP of a province, and each red bar on the adjacent left side to the blue bar represents the number of WM users located in that province. It is interesting to find that the top-three developed provinces that have the largest GDP almost have the greatest number of WM users. Meanwhile, although some low-GDP provinces have a greater number of WM users than some high-GDP provinces, the results show a positive correlation between the GDP and the number of WM users in each province.

Furthermore, we plot the correlation of GDP and number of WM users in Figure~\ref{fig:bian-usr-location}, and it is clear that there exists a linear correlation between them, which can be represented by a linear function $y = 0.0004493 \cdot x + 0.1751$.

\textbf{Implications}: The telecommunication sector in China has a hybrid and complex market, where high-end smart phones and outdated feature phones coexist; and multiple cellular/broadband data access methods of 2G, 3G, and 4G coexist. The higher GDP a province has, the more users in that province can afford to use a smart phone and pay a data access plan. Because WeChat is data-demanding and only available on smart phones, it is easy to understand why the GDP of a province has a linear correlation with the number of WM users in that province of China.




\section{Influence of WM Pages}~\label{sec:post}
In this section, we analyze the influence of WM pages by looking at four metrics: a page's life time~\cite{guo2009analyzing}, the number of page views before reposting, and the numbers of page views and infected users in a certain time period. Albeit the diffusion restrictions imposed by WM, businesses still wish their posted pages to be highly influential, e.g., to stay alive for a long time, to get reposted easily, to be viewed many times or by many users in a short time period.



\subsection{Page life time}


We define the \emph{life time} of a WM page as the time period from the time point of its first release in the WM network to the time point when it becomes \emph{inactive}. A WM page is \emph{inactive} when it has no infected users any more for at least one day.


%
%


We plot how many pages can stay alive in a certain period after the first release of these pages in Figure~\ref{fig:post-life}. We observe that the number of alive pages decline rapidly in the first day of the page release, while the decreasing rate becomes smaller thereafter. About 40\% pages in WM have a life time less than one day, due to the strong restrictions on WM page diffusions. However, as long as pages get through the first day, they are likely to have a long life time. Meanwhile, almost all pages become inactive 45 days after the page release.

\textbf{Implications}: The service provider tries to avoid information overload in WM; and meanwhile it launches it own advertising system in 2015 to deliver advertisements to WeChat users. Hence, it hopes that the diffusion of WM pages is restricted within a small community, and the pages become inactive in a short time period. As a result, even if a page is delivered to many users on its release day, it becomes inactive quickly as most infected users may not be willing to repost it (i.e., most users are not infectious), which reveals the temporal influence reduction of WM pages. This result is useful when businesses consider how to do marketing in the WM networks. It is recommended to release the WM pages to multiple communities of users, at different time points, in order to induce more infectious users and make more pages get through their first day.




\subsection{Number of page views before reposting}
\begin{figure}[ht]
\begin{minipage}[t]{0.45\linewidth}%
\centering
\includegraphics[width=1.5in]{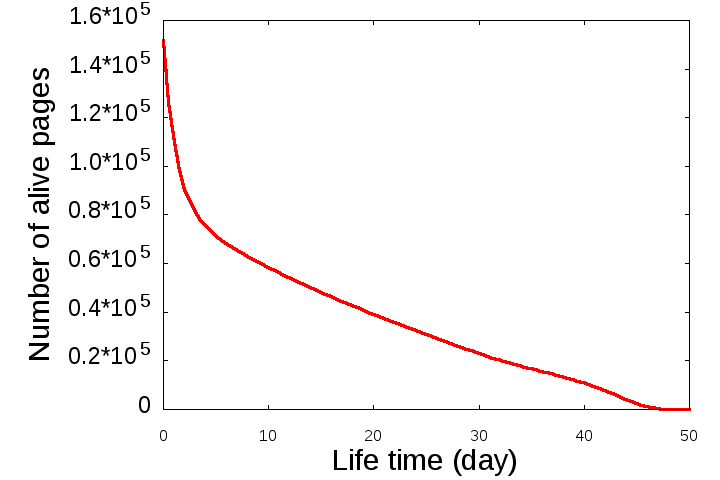}
\caption{The life time of pages.} \label{fig:post-life}
\end{minipage}
\hfill%
\begin{minipage}[t]{0.45\linewidth}%
\centering
\includegraphics[width=1.5in]{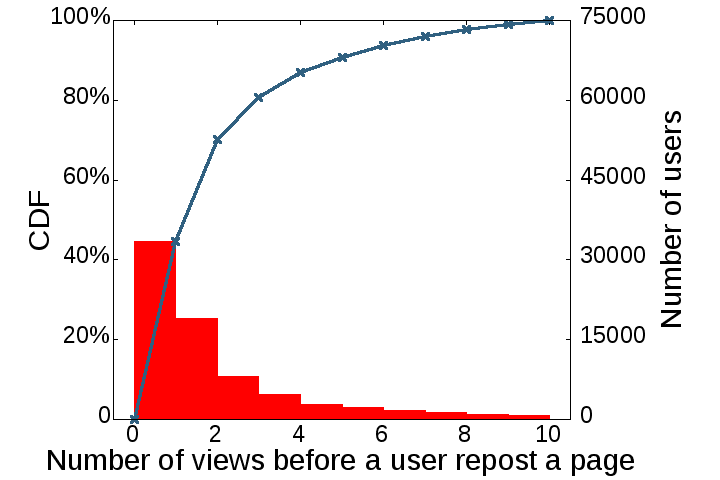}
\caption{The number of page views from a user before a page reposting.} \label{fig:view-before-post} %
\end{minipage}
\end{figure}

A highly influential page is expected to be reposted by many users. However, a user may have a hard time deciding whether to repost a page, as we found  users may view a page multiple times before he reposts it. To figure out the relationship between the number of page views and the user's intention of reposting a page, we take a look at how many times users view a WM page before they repost it, the results of which are given in Figure~\ref{fig:view-before-post}.

From the results, we observe that about 40\% of users repost the WM page the first time he views it. While the rest of users will not repost the WM page until they see it multiple times. In other words, users are more likely to repost a WM page if they see many of their friends have posted it. Meanwhile, we observe from the CDF that above 90\% of infectious users have viewed a page no more than 6 times before they repost it, which indicates that the critical page view number to stimulate a page reposting is 6.

\textbf{Implications}: According to the herd mentality, a user is likely to believe a page worth reposting, and then repost it, if he sees a page from his friends repeatedly. For promotions in WM, businesses should make their target customers infected multiple times; moreover, they should try to infect multiple users in a community where users are highly likely to be friends in WM, which makes an infected user believe the page is worth reposting.


\begin{figure}[ht]
\begin{minipage}[t]{0.45\linewidth}%
\centering
\includegraphics[width=1.5in]{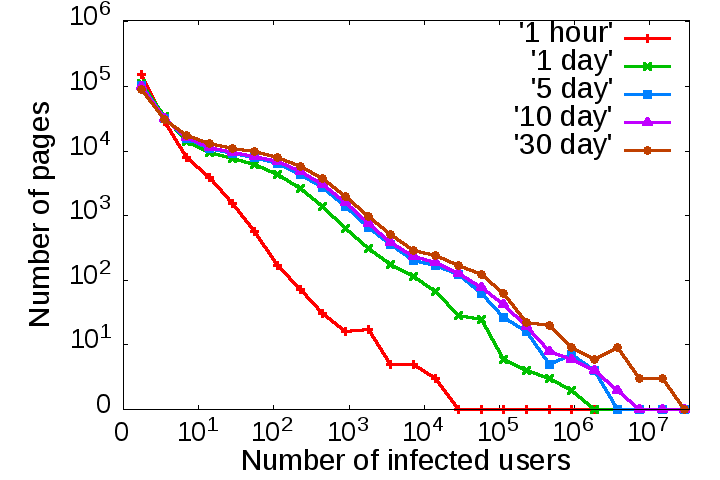}
\caption{The number of infected users by WM pages.} \label{fig:post-people-time}
\end{minipage}
\hfill%
\begin{minipage}[t]{0.45\linewidth}%
\centering
\includegraphics[width=1.5in]{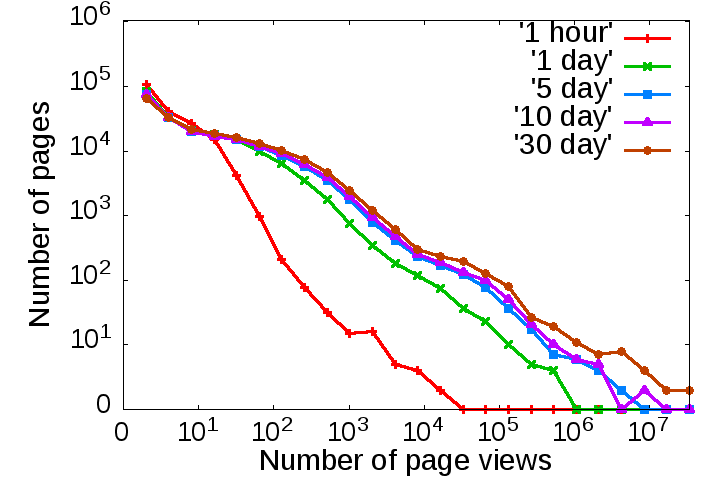}
\caption{The number of views for WM pages.} \label{fig:post-viewed-time}
\end{minipage}
\end{figure}

%

\subsection{Numbers of infected users and page views}
In this section, we examine the number of infected users and the number of page views of individual pages during different time periods, both of which are strong indicators of the page influence. Note that the second number is no smaller than the first one, as one user may view a page multiple times.

First, we choose five time periods after a page gets its first view: one hour, one day, five days, ten days and thirty days. For these time periods, we plot the number of infected users in Figure~\ref{fig:post-people-time}. We observe that every curve starts at a point with coordinates of $(3, 10^5)$, which implies that about $10^5$ pages can infect no more than three users in the first hour after getting their first page view.
We also notice that there exists a big gap between the curves of one hour and one day, which indicates that a large portion of users are infected by the pages in the first day after the first page view.

We also take a look at the number of page views, and plot the results in Figure~\ref{fig:post-viewed-time}. We find that the curve for each time period in Figure~\ref{fig:post-viewed-time} is slightly higher than the corresponding one in Figures~\ref{fig:post-people-time}, and the gap between the curves of one hour and one day exists in both figures. This result, again, verifies that the first day is important for a WM page to get popular.

\textbf{Implications}: For a given user, the pages posted by his friends are displayed to the user in a chronological order---the latest page is placed on the top. As time goes on, few users will bother to scroll down to view the page days before, and thus after five days the gap between curves becomes pretty small. Based on these results, we can conclude that it does not take a long time for a page in WM to become popular; meanwhile, it also takes a short time for a popular page to lose users' attention and page views. Hence, for businesses, it is important to make their pages well spread on the first day, and trying to allure more users to repost is also a trick to have a long life time for their pages (which is beyond the scope of this paper).

\section{User Influence}~\label{sec:user}
In this section, we investigate a user's influence by studying the individual users' behaviors of page view and reposting.

%
%
%

\subsection{Number of viewers for individual pages}
One's influence can be somewhat reflected by the average number of viewers for each of his individual pages. 

\begin{figure}[ht]
\begin{minipage}[t]{0.45\linewidth}%
\centering
   \includegraphics[width=1.5in]{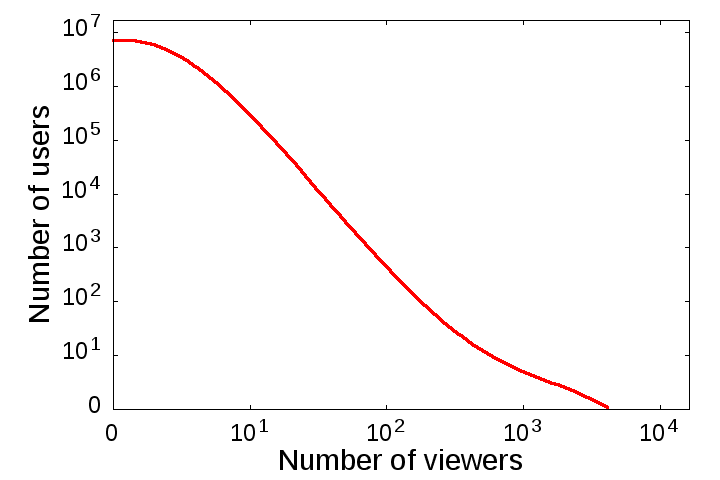}
\caption{Number of viewers for individual pages.}
\label{fig:avg-range}
\end{minipage}
\hfill%
\begin{minipage}[t]{0.45\linewidth}%
\centering
   \includegraphics[width=\columnwidth]{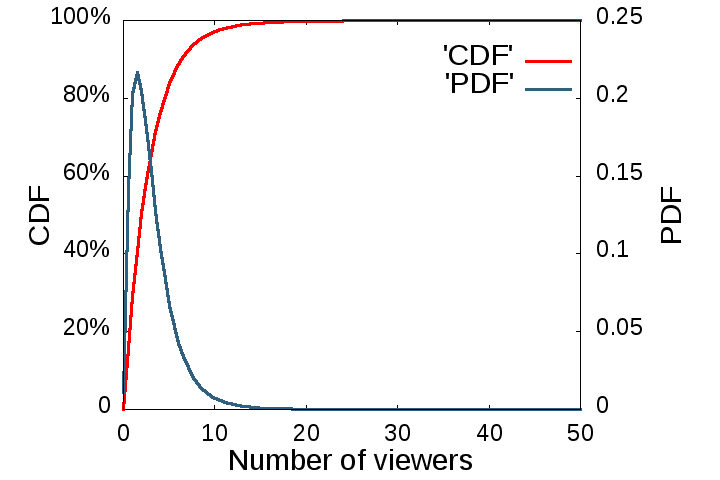}
\caption{CDF and PDF for the number of viewers for individual pages.}
\label{fig:avg-range-cdf}
\end{minipage}
\end{figure}

In Figure~\ref{fig:avg-range}, the horizontal axis (using a log scale) denotes the average number of viewers for one's individual pages, and the vertical axis (using a log scale) denotes the number of eligible users that have a certain number of viewers. We can observe that the logarithm of the number of eligible users for a given number of viewers declines with the increase of the average number of viewers.
It is also noteworthy that there is no user whose average number of viewers is greater than $5,000$, since WM places a limit of $5,000$ over the maximum number of friends for a user.

Figure~\ref{fig:avg-range-cdf} illustrates the corresponding CDF and PDF. We observe from the CDF curve that $80\%$ of users have the average number of viewers less than $5$, and more than $92\%$ of users have the average number of viewers less than $10$. The PDF curve has a peak when the average number of viewers is 3, which means that a user is mostly likely to have three viewers on average for each of his individual pages.

\textbf{Implications}: These results confirm that WM is successful in restricting one's influence, as the vast majority of users have a very small number of  viewers for individual pages. This suggests businesses discard the conventional marketing strategy of recruiting KOL (key opinion leader) users who have millions of fans/followers in conventional social networks (e.g., Twitter, or Weibo).


\subsection{Number of identified friends}\label{sec:infludeg}

Yet another indicator of one's influence in our study is the \emph{number of identified friends}. Users~$i$ and $j$ are \emph{identified friends}, if the record in our dataset shows either user~$i$ views user~$j$'s posted pages, or user~$j$ views user~$i$'s posted pages. Note that the number of identified friends of user~$i$ may be smaller than his true number of friends, as the dataset only collects statistics of the diffusion graph instead of the underlying WM network. 

\begin{figure}[ht]
\begin{minipage}[t]{0.45\linewidth}%
\centering
   \includegraphics[width=\columnwidth]{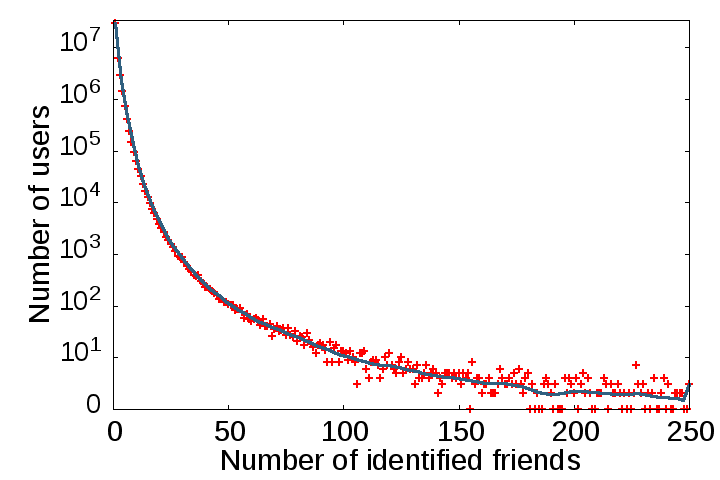}
\caption{Number of identified friends}
\label{fig:range-degree}
\end{minipage}
\hfill%
\begin{minipage}[t]{0.45\linewidth}%
\centering
   \includegraphics[width=\columnwidth]{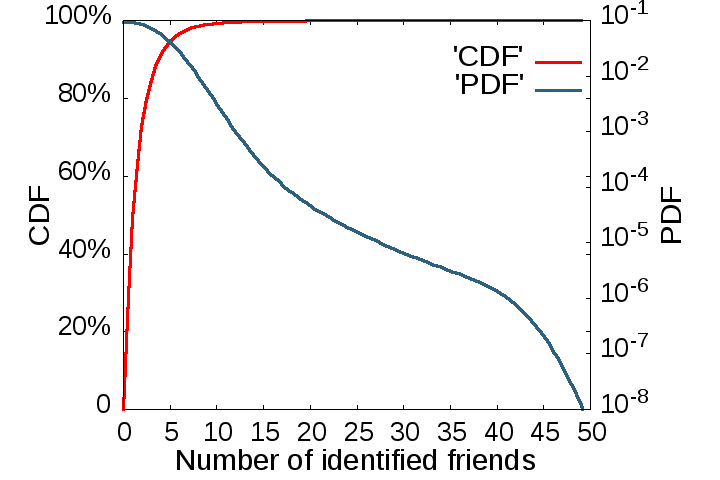}
\caption{CDF and PDF for the number of identified friends.}
\label{fig:range-degree-cdf}
\end{minipage}
\end{figure}

The distribution of number of identified friends for individual users is presented in Figure~\ref{fig:range-degree}.
The $y$-axis coordinate of each red point in the figure indicates the logarithm of the number of users that have a certain number of identified friends (represented by the $x$-axis coordinate).  We notice that some red points reside on the horizontal axis, which implies that no user has such a number of identified friends according to the statistics in our dataset. The blue curve is the regression curve for the red scattered points. The number of users is noted to decline super-exponentially with an increase in the number of identified friends.


The CDF and PDF curves of the number of identified friends are presented in Figure~\ref{fig:range-degree-cdf}. We observe that more than $92\%$ of users have a number of identified friends less than $5$. Meanwhile, the PDF curves indicates that the probability that a user has a large number of identified friends is small.

\textbf{Implications}: Most users cannot have a large number of identified friends according to the diffusion graph, and thus the number of friends is no longer an indicator of one's influence in the information diffusion process. Instead, businesses need to refer to other criteria for judging the user influence.

%



\subsection{User similarity}

%

In this section, we define two types of similarity between users: the influence similarity, and the location similarity.

\begin{figure}[ht]
\begin{minipage}[t]{0.45\linewidth}%
\centering
   \includegraphics[width=\columnwidth]{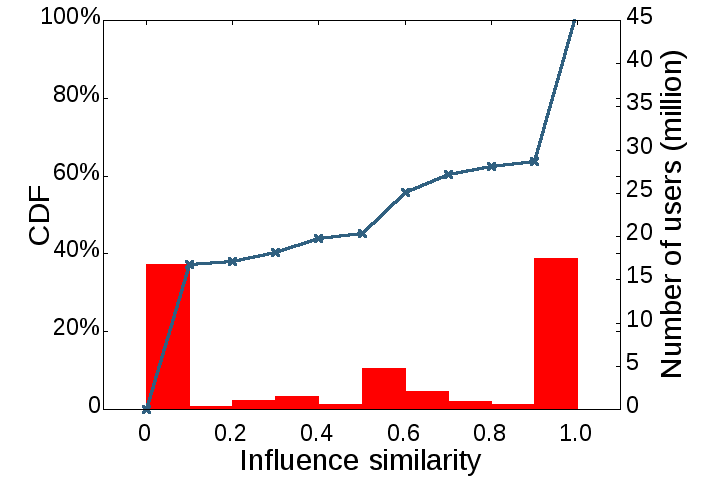}
\caption{Influence similarity (The curve representing the CDF and bars representing the number of users).}
\label{fig:simi}
\end{minipage}
\hfill%
\begin{minipage}[t]{0.45\linewidth}%
\centering
    \includegraphics[width=\columnwidth]{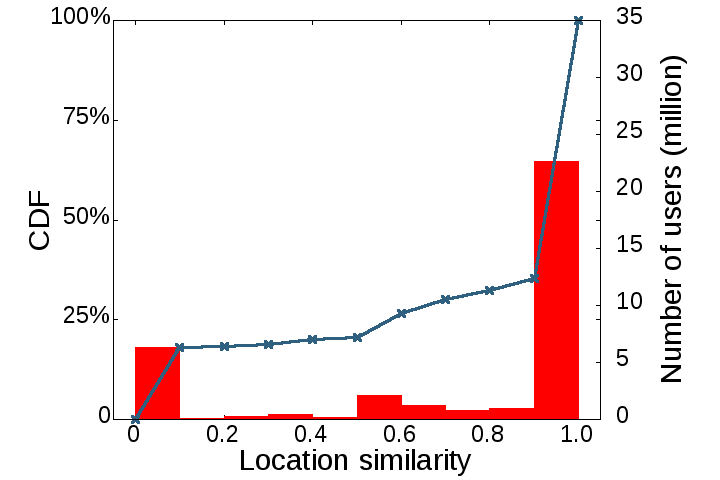}
\caption{Location similarity (The curve representing the CDF and bars representing the number of users).} \label{fig:bian-loc-similarity}
\end{minipage}
\end{figure}

The influence similarity is used to compare two users to see if they have similar number of identified friends. Let $\mathrm{F}(\cdot)$ denotes the number of identified friends of a user,
We consider users~$i$ and $j$ are \emph{similar in the number of identified friends} if
\[
\frac{1}{2}< \frac{\mathrm{F}(i)}{\mathrm{F}(j)}<2,
\]
A user $i$'s influence similarity is defined as the proportion of his identified friends who are similar to user~$i$ in terms of the number of identified friends.
Figure~\ref{fig:simi} illustrates the distribution of users' influence similarity, where we observe a dichotomous distribution. Approximately $25\%$ of users (the leftmost bar representing about 18 million of users) have a influence similarity value less than $0.1$, while another $25\%$ users (the rightmost bar representing about 18 million of users) have a value greater than $0.9$.
Why do such two groups of users dominate in the distribution of influence similarity? We inspect the statistics of these two group of users, and find that the users in the two groups usually have a small number (one or two) of identified friends. A small number of identified friends could lead to extreme values of the influence similarity of a user: in the first group, it could be very low when one's friends have a quite different number of identified friends; and in the second group, it could be very high when his friends have a similar number of identified friends.


Then, we consider the location similarity: users~\emph{i} and~\emph{j} are viewed \emph{similar in location} if they are located in the same province. A user $i$'s location similarity is defined as the proportion of his identified friends who are similar in location to user~$i$.
We observe that in Figure~\ref{fig:bian-loc-similarity}, about 65\% users (the rightmost bar representing about 23 million of users) have a location similarity value above 90\%. This result shows that most users in WM prefer to make friends with those who has a similar location. We also notice that about 18\% users' location similarity is below 10\% (the leftmost bar). The reason for the leftmost bar in this distribution is similar to that for the distribution of influence similarity: when a user has a small number (one or two) of identified friends, it is likely that his friends have different locations, leading to a low location similarity.

\textbf{Implications}: From the results, we learn that a user with a small number of identified friends is not ``similar'' to his friends. Such a user is not a good candidate for spreading information in WM, because it is difficult to stimulate his friends to repost his pages if his friends find they are quite different~\cite{mcpherson2001birds}.


\subsection{Page view statistics by a user}
\begin{figure}[ht]
\begin{minipage}[t]{0.45\linewidth}%
\centering
   \includegraphics[width=\columnwidth]{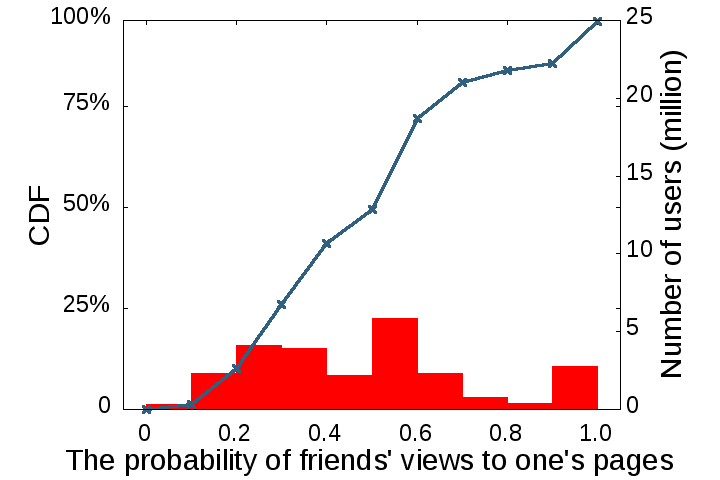}
\caption{The probability of friends' views to one's pages.}
\label{fig:poss}
\end{minipage}
\hfill%
\begin{minipage}[t]{0.45\linewidth}%
 \centering
   \includegraphics[width=\columnwidth]{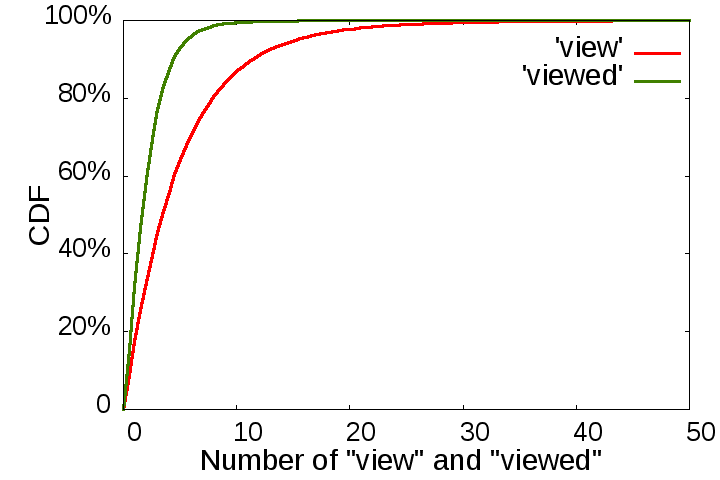}
\caption{CDFs of numbers of ``view'' and ``viewed'', per day.}
\label{fig:view}
\end{minipage}
\end{figure}

Now we discuss the view statistics of an individual user's posts. We consider the probability that a user's friends viewed his posted pages. Let $A(i)$ be the set of all posted pages by user $i$. For a page $a\in A(i)$, let $V(a)$ be the set of friends that have viewed this page $a$. Thus the total number of friends that have ever viewed user $i$'s pages is given by $|\bigcup_{a\in A(i)} V(a)|$.
The probability that a page $a$ is viewed by user $i$'s friends is
\[
\frac{|V(a)|}{|\bigcup_{a\in A(i)} V(a)|}.
\]
The probability that user $i$'s post is viewed by its friends is given by
\[
\frac{1}{|A(i)|}\cdot \frac{|V(a)|}{|\bigcup_{a\in A(i)} V(a)|}.
\]
Figure~\ref{fig:poss} shows the distribution of the probability of friends' views. The CDF is noted to increase from $0$ to $1$ in an approximately linear manner, which implies that the distribution is rather uniform.

We define the number of ``daily view'' for a user as the number of times that a user views others' pages per day; meanwhile, we define the number of ``daily viewed'' for a user, as well as the number of times that a user's pages are viewed by others per day. Figure~\ref{fig:view} shows the CDFs of these two numbers (the red and green curves). From Figure~\ref{fig:view}, we can clearly see the results that on average, a user's number of ``view'' (less than 30 with a probability of 0.99) is greater than the number of ``viewed'' (less than 10 with a probability of 0.99) per day.

\textbf{Implications}: These results show that a user's intention of viewing others' posted pages in WM is not strong, and moreover the friends' intention of viewing his pages is even more weakened. The key factors that stimulate users to view a page are merely the icon figure and the title of the shared page (as is shown in Figure~\ref{fig:like_comment}). Hence, many businesses have tried playing word games in the title of pages to attract more page views, which further deteriorates the user experience in WM.


%



\section{Conclusions}~\label{sec:conclusoin}
In this paper, we present a measurement study on three aspects of the information diffusion process in the WeChat Moments (WM) network, based on a dataset of more than 160,000 WM pages created by businesses in 2015 (with an average number of 10 million page views every day by 40 million users). Specifically, we analyze the overall diffusion graph in WM, examine the influence of individual WM pages, and evaluate a user's influence under the restrictions in the WM networks. Moreover, we provide discussions on the experiment results and their implicit connections to macro-economic factors (e.g., GDP of a user's province), and development of marketing strategies for better publicity. To the best of our knowledge, this is the first measurement study on the information diffusion in WeChat Moments networks.

\section{Acknowledgements}~\label{sec:acknowledgements}
This work was sponsored by National Natural Science Foundation of China (NSFC) under grant number 61572051 and 61401169.
%
\balance

\bibliographystyle{abbrv}
\bibliography{reference-list}

\end{document}